\documentclass[prd,aps,showpacs,nofootinbib]{revtex4}%
\usepackage{amsmath}
\usepackage{amsfonts}
\usepackage{amssymb}
\usepackage{bm}%
\usepackage{graphicx}

\begin{document}
\title{ Gauge propagator and physical consistency of the CPT-even part of the
Standard Model Extension}
\author{Rodolfo Casana$^{a}$, Manoel M. Ferreira Jr$^{a}$, Adalto R. Gomes$^{b}$,
Paulo R. D. Pinheiro$^{a}$}
\affiliation{$^{a}$Departamento de F\'{\i}sica, Universidade Federal do Maranh\~{a}o
(UFMA), Campus Universit\'{a}rio do Bacanga, S\~{a}o Lu\'{\i}s - MA,
65085-580, Brasil.}
\affiliation{$^{b}$Departamento de F\' isica, Instituto Federal de Educa\c c\~ao, Ci\^encia
e Tecnologia do Maranh\~ao (IFMA), 65025-001, S\~ao Lu\'\i s, Maranh\~ao, Brazil}

\begin{abstract}
In this work, we explicitly evaluate the gauge propagator of the Maxwell
theory supplemented by the CPT-even term of the SME. First, we specialize our
evaluation for the parity-odd sector of the tensor $W_{\mu\nu\rho\sigma}%
$,\ using a parametrization that retains only the three nonbirefringent
coefficients.\ From the poles of the propagator, it is shown that \ physical
modes of\textbf{\ }this electrodynamics are stable, non-causal and unitary. In
the sequel, we carry out the parity-even gauge propagator using a
parametrization that allows to work with only the isotropic nonbirefringent
element. In this case, we show that the\textbf{\ }physical modes of the
parity-even sector of the tensor $W$ are causal, stable and unitary for a
limited range of the isotropic coefficient.

\end{abstract}

\maketitle

\section{Introduction}

Lorentz symmetry violation has been intensively investigated in the latest
years. Most investigations have been performed into the body of the Standard
Model Extension (SME), developed by Colladay \& Kostelecky \cite{Colladay},
using the idea of spontaneous breaking of Lorentz symmetry in the context of
string theory \cite{Samuel}. The SME incorporates terms of Lorentz invariance
violation (LIV) in all sectors of interaction and has been studied in many
aspects \cite{General}. The investigations in the context of the SME concern
mainly the fermion sector \cite{Fermion,Lehnert} (also involving CPT tests
\cite{Tests}) and the gauge sector. The gauge sector of the SME is composed of
a CPT-odd and a CPT-even term. The CPT-odd part of SME is represented by the
Carroll-Field-Jackiw (CFJ) electrodynamics \cite{Jackiw}, whose properties
have been examined in several distinct aspects. Its consistency was addressed
in Refs. \cite{Adam,Soldati}, revealing a consistent (causal, stable and
unitary) model only for a space-like background. A version of this model
incorporating the Higgs sector was developed in Ref. \cite{Higgs}, in which
its gauge propagator was carried out and its consistency was analyzed. The
dimensionally reduced version of this theory was developed and examined in
Ref. \cite{Belich}, in which it was demonstrated to be causal, stable and
unitary. The stationary classical solutions for the CFJ electrodynamics were
attained in Ref. \cite{Casana}, whereas the vacuum emission of Cerenkov
radiation induced by the CFJ term was studied in Refs. \cite{Cerenkov1}. Works
discussing the finite temperature of the SME electrodynamics and its relation
with the black body radiation \cite{CBR,winder,CBR2} should also be mentioned.
Moreover, the issue about the radiative generation of the CFJ term has
engendered many papers as well \cite{Radio}.

Recently, the CPT-even sector of SME has been also much investigated
\cite{KM1,KM2,KM3,Bailey,Kostelec}, mainly in connection with issues able to
yield good bounds on the LIV coefficients. The study of Cherenkov radiation
\cite{Cherenkov2} and the absence of emission of Cherenkov radiation by UHECR
(ultrahigh energy cosmic rays) \cite{Klink2,Klink3} has been a point of strong
interest in latest years, as well as photon-fermion vertex interactions
yielding new bounds on the LIV coefficients \cite{Interac1,Interac2,Interac3}.
The classical solutions for this electrodynamics were also discussed both for
the parity-odd \cite{Manojr1} and parity-even sectors\cite{Paulo}. However,
the question concerning the evaluation of the gauge propagator and the
consistency of this electrodynamics has not been investigated so far. This is
the main purpose of the present work.

We begin presenting some general features of the CPT-even part of the SME
gauge sector, described by the following Lagrangian:
\begin{equation}
\mathcal{L}=-\frac{1}{4}F_{\alpha\nu}F^{\alpha\nu}-\frac{1}{4}W_{\alpha\nu
\rho\varphi}F^{\alpha\nu}F^{\rho\varphi},\label{L1}%
\end{equation}
where $F_{\alpha\nu}$ is the electromagnetic field tensor, $W_{\alpha\nu
\rho\varphi}$ is a renormalizable and dimensionless coupling, responsible for
Lorentz violation. The tensor $W_{\alpha\nu\rho\varphi}$ has the same
symmetries as the Riemann tensor
\begin{equation}
W_{\alpha\nu\rho\varphi}=-W_{\nu\alpha\rho\varphi}\text{ },W_{\alpha\nu
\rho\varphi}=-W_{\alpha\nu\varphi\rho},~W_{\alpha\nu\rho\varphi}%
=W_{\rho\varphi\alpha\nu},
\end{equation}
and a double null trace, $W_{\text{ \ \ \ }\rho\varphi}^{\rho\varphi}=0$. As
the tensor $W_{\alpha\nu\rho\varphi}$ has in principle 19 components, it is
necessary to use some parametrization to turn feasible the study of its
effects on the Maxwell theory. A very useful parametrization is the one
presented in Refs. \cite{KM1,KM2} which encloses these 19 components in a
parity-even and a parity-odd subsectors, represented by the matrices
$\widetilde{\kappa}_{e}$ and $\widetilde{\kappa}_{o}$, respectively,
\begin{align}
\left(  \widetilde{\kappa}_{e+}\right)  ^{jk} &  =\frac{1}{2}(\kappa
_{DE}+\kappa_{HB})^{jk},~~\left(  \widetilde{\kappa}_{e-}\right)  ^{jk}%
=\frac{1}{2}(\kappa_{DE}-\kappa_{HB})^{jk}-\frac{1}{3}\delta^{jk}(\kappa
_{DE})^{ii},~~\kappa_{\text{tr}}=\frac{1}{3}\text{tr}(\kappa_{DE}),~~\\
\left(  \widetilde{\kappa}_{o+}\right)  ^{jk} &  =\frac{1}{2}(\kappa
_{DB}+\kappa_{HE})^{jk},\text{ ~~}\left(  \widetilde{\kappa}_{o-}\right)
^{jk}=\frac{1}{2}(\kappa_{DB}-\kappa_{HE})^{jk}~~.
\end{align}
The $3\times3$ matrices $\kappa_{DE},\kappa_{HB},\kappa_{DB},\kappa_{HE} $ are
given as:
\begin{align}
\left(  \kappa_{DE}\right)  ^{jk} &  =-2W^{0j0k},\text{ }\left(  \kappa
_{HB}\right)  ^{jk}=\frac{1}{2}\epsilon^{jpq}\epsilon^{klm}W^{pqlm}%
,\label{P1}\\
\text{ }\left(  \kappa_{DB}\right)  ^{jk} &  =-\left(  \kappa_{HE}\right)
^{kj}=\epsilon^{kpq}W^{0jpq}.\label{P2}%
\end{align}
The matrices $\kappa_{DE},\kappa_{HB}$ contain together 11 independent
components while $\kappa_{DB},\kappa_{HE}$ possess together 8 components,
which sums the 19 independent elements of the tensor $W_{\alpha\nu\rho\varphi
}$.

In terms of the matrices $\kappa_{DE},$ $\kappa_{HB},\kappa_{DB},\kappa_{HE}$,
the Lagrangian (\ref{L1}) is read as
\begin{equation}
\mathcal{L}=\frac{1}{2}\left(  \mathbf{E}^{2}-\mathbf{B}^{2}\right)  +\frac
{1}{2}\mathbf{E}\cdot\left(  \kappa_{DE}\right)  \cdot\mathbf{E}-\frac{1}%
{2}\mathbf{B}\cdot\left(  \kappa_{HB}\right)  \cdot\mathbf{B}+\mathbf{E}%
\cdot\left(  \kappa_{DB}\right)  \cdot\mathbf{B}-\rho A_{0}+\mathbf{j\cdot
A},\label{L1B}%
\end{equation}
while in terms of the matrices $\left(  \widetilde{\kappa}_{e+}\right)  ,$
$\left(  \widetilde{\kappa}_{e-}\right)  ,$ $\left(  \widetilde{\kappa}%
_{o+}\right)  $, $\left(  \widetilde{\kappa}_{o-}\right)  $, and the
coefficient $\left(  \kappa_{\text{tr}}\right)  $, the Lagrangian is
\begin{align}
\mathcal{L} &  =\frac{1}{2}\left[  \left(  1+\kappa_{\text{tr}}\right)
\mathbf{E}^{2}-\left(  1-\kappa_{\text{tr}}\right)  \mathbf{B}^{2}\right]
+\frac{1}{2}\mathbf{E}\cdot\left(  \widetilde{\kappa}_{e+}+\widetilde{\kappa
}_{e-}\right)  \cdot\mathbf{E}\nonumber\\
&  -\frac{1}{2}\mathbf{B}\cdot\left(  \widetilde{\kappa}_{e+}-\widetilde
{\kappa}_{e-}\right)  \cdot\mathbf{B}+\mathbf{E}\cdot\left(  \widetilde
{\kappa}_{o+}+\widetilde{\kappa}_{o-}\right)  \cdot\mathbf{B}-\rho
A_{0}+\mathbf{j\cdot A}.\label{L1C}%
\end{align}

The investigation about the properties of Lagrangian (\ref{L1}) were initiated
by Kostelecky \& Mewes in Ref. \cite{KM1}, where it was stipulated the
existence of ten linearly independent combinations of the components of
$W_{\alpha\nu\rho\varphi}$ sensitive to birefringence. These elements are
contained in the matrices $\widetilde{\kappa}_{e+}$ and $\widetilde{\kappa
}_{o-}.$ Using high-quality spectropolarimetry data of cosmological sources,
an upper bound as stringent as $10^{-32}$ was imposed on these birefringent
LIV parameters. In Ref. \cite{KM2}, these authors have confirmed these results
and stated new bounds on the nonbirefringent components using microwave
cavities experiments. Recently, these authors have constrained some
birefringent coefficients to the level of one part in $10^{37}$ using linear
polarization data of gamma rays emitted from cosmological sources \cite{KM3}.
On the other hand, general properties of the electrodynamics of Lagrangian
(\ref{L1}) in continuous media were discussed in Ref. \cite{Bailey}, where
some stationary solutions were also obtained. More recently, there appeared
some works coupling the modified Maxwell electrodynamics of Eq. (\ref{L1}) to
the Dirac sector \cite{Klink2}. As a result, new bounds on the nine
nonbirefringent components of the tensor $W_{\alpha\nu\rho\varphi}$ (including
the six ones of the parity even sector) were obtained from the absence of
vacuum Cherenkov radiation associated with ultrahigh-energy cosmic rays
(UHECRs). In Ref. \cite{Klink3} the LIV tensor $W_{\alpha\nu\rho\varphi}$ was
reduced to only one element, the trace of the parity-even sector. The coupling
of this electrodynamics with the Dirac sector generates a modified quantum
electrodynamics, in which it was evaluated the decay rates for processes
involving Cherenkov radiation. A two-sided bound was then stated for the trace
parameter. Recently, the issue concerning the stationary solutions for the
electrodynamics of Eq. (\ref{L1}) was suitably addressed by means of the Green
method. Starting from the Maxwell and wave equations, classical solutions were
found for the parity-odd sector \cite{Manojr1} and for the parity-even sector
\cite{Paulo}, with upper bounds as good as one part in $10^{20}$.

The purpose of the present work is to evaluate the gauge propagator of
Lagrangian (\ref{L1}), addressing both the parity-odd and parity-even sectors
of the tensor $W_{\alpha\nu\rho\varphi}.$ We then use the pole structure of
the propagator for studying the stability, causality, unitarity (consistency)
and the dispersion relations of this model. We first carry out the propagator
for the parity-odd sector reduced to the only three nonbirefringent
components. Such propagator is written as a $4\times4$ matrix. The pole
analysis shows that this sector is stable, non-causal and unitary. In the
sequel, we evaluate the propagator of the parity-even sector, only represented
by the trace isotropic component. The pole analysis showed that this sector is
stable, causal and unitary for $0\leq\kappa_{\text{tr}}<1$. In both sectors,
it was observed the presence of a second order pole, $p^{2}=0$, whose residue
is gauge dependent (see Appendix \ref{apendice1}). So, it does not contribute
to the S-matrix, being non-physical.

\section{The gauge propagator}

To compute the propagator of the gauge field described by the Lagrangian
density (\ref{L1}), we define the generating functional of the Green's
function which, in the Lorentz gauge, is given by
\begin{equation}
Z\left[  J^{\mu}\right]  =N\int\mathcal{D}A_{\mu}\!\!\exp\left\{
i\int\!\!d^{4}x\left(  -\frac{1}{4}F_{\mu\nu}F^{\mu\nu}-\frac{1}{4}%
W^{\alpha\nu\rho\varphi}F_{\alpha\nu}F_{\rho\varphi}-\frac{1}{2\xi}\left(
\partial_{\mu}A^{\mu}\right)  ^{2}+A_{\mu}J^{\mu}\right)  \right\}  ,
\end{equation}
where $N$ is a normalization factor satisfying $Z\left[  0\right]  =1$ and
$\xi$ is the gauge-fixing parameter. After some integration by parts, we get%
\begin{equation}
Z\left[  J^{\mu}\right]  =\int\mathcal{D}A_{\mu}\!\!\exp\left\{
i\int\!\!d^{4}x~\left(  \frac{1}{2}A_{\mu}D^{\mu\nu}A_{\nu}+A_{\mu}J^{\mu
}\right)  \right\}  ,
\end{equation}
with $D^{\mu\nu}$ being a second order operator defined as
\begin{equation}
D^{\mu\nu}=\square g^{\mu\nu}+\left(  \frac{1}{\xi}-1\right)  \partial^{\mu
}\partial^{\nu}-S^{\mu\nu},
\end{equation}
whereas $g^{\mu\nu}=(+,---)$\ is the metric tensor. We have also defined the
symmetric operator $S^{\mu\nu}$
\begin{equation}
S^{\mu\nu}=2W^{\mu\alpha\beta\nu}\partial_{\alpha}\partial_{\beta}=S^{\nu\mu
}.\label{S1}%
\end{equation}

Performing the gauge field integration, the generating functional becomes
\begin{equation}
Z\left[  J^{\mu}\right]  =\exp\left\{  -\frac{i}{2}\int\!\!dxdyJ^{\mu}\left(
x\right)  \Delta_{\mu\nu}\left(  x-y\right)  J^{\nu}\left(  y\right)
\right\}  ,
\end{equation}
where $\Delta_{\mu\nu}\left(  x-y\right)  $ is the Green's function, given as%
\begin{equation}
D^{\mu\beta}\Delta_{\beta\nu}\left(  x-y\right)  =\delta^{\mu}{}_{\!\nu}%
\delta\left(  x-y\right)  .
\end{equation}

The propagator of the gauge field is found to be%
\begin{equation}
\left\langle 0\left\vert TA_{\mu}\left(  x\right)  A_{\nu}\left(  y\right)
\right\vert 0\right\rangle =i\Delta_{\mu\nu}\left(  x-y\right)  .\label{propA}%
\end{equation}

Now, we compute the Green's functions in the Feynman gauge, $\xi=1$, which
implies $D^{\mu\nu}=\square g^{\mu\nu}-S^{\mu\nu}.$ The Green's function then
satisfy%
\begin{equation}
\left(  \square g^{\mu\beta}-S^{\mu\beta}\right)  \Delta_{\beta\nu}\left(
x-y\right)  =\delta^{\mu}{}_{\!\nu}\delta\left(  x-y\right)  .\label{PE2}%
\end{equation}

In the Fourier representation, we have
\begin{equation}
\delta\left(  x-y\right)  =\int\frac{\!\!dp}{\left(  2\pi\right)  ^{4}%
}e^{-ip\cdot\left(  x-y\right)  }~,~\ \ \Delta_{\beta\nu}\left(  x-y\right)
=\int\frac{\!\!dp}{\left(  2\pi\right)  ^{4}}\ \widetilde{\Delta}_{\beta\nu
}\left(  p\right)  e^{-ip\cdot\left(  x-y\right)  },
\end{equation}
with%
\begin{align}
\widetilde{D}^{\mu\beta}  &  =-\left(  p^{2}g^{\mu\beta}-\tilde{S}^{\mu\beta
}\right)  ,\\
\tilde{S}^{\mu\nu}  &  =2W^{\mu\alpha\beta\nu}p_{\alpha}p_{\beta}=\tilde
{S}^{\nu\mu}.\label{S01}%
\end{align}
The expression (\ref{PE2}) is then written as%
\begin{equation}
-\left(  p^{2}g^{\mu\beta}-\tilde{S}^{\mu\beta}\right)  \widetilde{\Delta
}_{\beta\nu}\left(  p\right)  =\delta^{\mu}{}_{\!\nu}.\label{PE3}%
\end{equation}
The gauge propagator will be evaluated by inverting this tensor expression
once some specialization is assumed for the tensor $\tilde{S}^{\mu\beta}.$
This is the objective of the next section.

\subsection{The gauge propagator for the parity-odd sector}

In this section, we are interested in evaluating the propagator for Maxwell
electrodynamics supplemented by the parity-odd sector of the tensor
$W^{\alpha\nu\rho\varphi},$ following the same parametrization adopted in
Refs. \cite{Manojr1},\cite{Kob}. It consists in taking as null the parity-even
sector ($\kappa_{DE}=\kappa_{HB}=0$ or $\widetilde{\kappa}_{e+}=\widetilde
{\kappa}_{e-}=\kappa_{\text{tr}}=0)$ and in retaining only three components of
the parity-odd sector (the ones supposed not constrained by birefringence).
Such parametrization is imposed by the conditions $(\kappa_{DB})=-\left(
\kappa_{HE}\right)  ^{T}$ and $\kappa_{DB}=\kappa_{HE}$, which turns the
$3\times3$ matrix $\kappa_{DB}=\kappa_{HE}$ anti-symmetric and justifies the
existence of sole three non-null elements. Thus we have $\left(
\widetilde{\kappa}_{o+}\right)  ^{jk}=(\kappa_{DB})^{jk}$. These three non
vanishing elements belong to the matrix  $\kappa_{DB}\ $and are written in
terms of the components of a three-vector $\boldsymbol{\kappa}$ at the form:
\begin{equation}
\kappa^{j}=\frac{1}{2}\epsilon^{jpq}\left(  \kappa_{DB}\right)  ^{pq}%
~.\label{def2}%
\end{equation}
These components are taken as nonbirefringent in Refs.
\cite{KM1,KM2,KM3,Kostelec,Manojr1,Kob} and, the non-null elements of the
tensor $W^{\mu\alpha\beta\nu}$ are given in terms of the $\mathbf{\kappa}$
vector by
\begin{equation}
W^{0ijl}=\frac{1}{2}\left[  \kappa^{j}\delta^{il}-\kappa^{l}\delta
^{ij}\right]  .\label{W-odd}%
\end{equation}
For this situation, the Lagrangian (\ref{L1C}) is reduced to the form
\begin{equation}
\mathcal{L}=\frac{1}{2}\left(  \mathbf{E}^{2}-\mathbf{B}^{2}\right)
+\boldsymbol{\kappa}\cdot(\mathbf{E}\times\mathbf{B})-\rho A_{0}%
+\mathbf{j\cdot A}.
\end{equation}
\bigskip

The more usual way for the evaluation of the propagator (\ref{propA}), in
momentum space, would be by means of the definition of a closed algebra
involving the operators $\ L^{\mu\nu}=p^{\mu}p^{\nu}/p^{2},$ $T^{\mu\nu
}=g^{\mu\nu}-L^{\mu\nu},\tilde{S}^{\mu\nu}=2W^{\mu\alpha\beta\nu}p_{a}%
p_{\beta},R^{\mu\nu}=\tilde{S}^{\mu\rho}\tilde{S}_{\rho}{}^{\nu}.$
Unfortunately, the search for a closed algebra involving a combination of
these operators has been unsuccessful. Given the impossibility of finding a
closed algebra that yields the attainment of a tensor form for the propagator,
we have adopted the strategy of writing the full matrix that represents the
operator $\widetilde{D}^{\mu\nu}=-\left(  p^{2}g^{\mu\beta}-\tilde{S}%
^{\mu\beta}\right)  ,$ inverting it in the sequel. Thus, we begin evaluating
all the components of the operator $\tilde{S}^{\mu\nu}$, given by Eq.
(\ref{S01}) by using the parametrization (\ref{W-odd}), which can be properly
grouped in the $4\times4$ matrix:
\begin{equation}
\tilde{S}^{\mu\nu}=\left(
\begin{array}
[c]{ccccccc}%
0 &  & \kappa_{1}\mathbf{p}^{2}-p_{1}A &  & \kappa_{2}\mathbf{p}^{2}-p_{2}A &
& \kappa_{3}\mathbf{p}^{2}-p_{3}A\\[0.2cm]%
\kappa_{1}\mathbf{p}^{2}-p_{1}A &  & 2p_{0}\left(  A-\kappa_{1}p_{1}\right)
&  & -p_{0}\left(  \kappa_{1}p_{2}+\kappa_{2}p_{1}\right)   &  & -p_{0}\left(
\kappa_{3}p_{1}+\kappa_{1}p_{3}\right)  \\[0.2cm]%
\kappa_{2}\mathbf{p}^{2}-p_{2}A &  & -p_{0}\left(  \kappa_{1}p_{2}+\kappa
_{2}p_{1}\right)   &  & 2p_{0}\left(  A-\kappa_{2}p_{2}\right)   &  &
-p_{0}\left(  \kappa_{2}p_{3}+\kappa_{3}p_{2}\right)  \\[0.2cm]%
\kappa_{3}\mathbf{p}^{2}-p_{3}A &  & -p_{0}\left(  \kappa_{3}p_{1}+\kappa
_{1}p_{3}\right)   &  & -p_{0}\left(  \kappa_{2}p_{3}+\kappa_{3}p_{2}\right)
&  & 2p_{0}\left(  A-\kappa_{3}p_{3}\right)
\end{array}
\right)  ,
\end{equation}
where $A=\left(  \boldsymbol{\kappa}\cdot\mathbf{p}\right)  .$ We can now show
some results of interest:
\begin{align}
S^{\alpha}{}_{\alpha} &  =S^{0}{}_{0}+S^{1}{}_{1}+S^{2}{}_{2}+S^{3}{}%
_{3}=-4p_{0}(\boldsymbol{\kappa}\cdot\mathbf{p}),\label{ST1}\\[0.2cm]
S^{\mu\rho}S_{\mu\rho}{} &  =-2\boldsymbol{\kappa}^{2}\mathbf{p}^{4}+2\left(
\boldsymbol{\kappa}\cdot\mathbf{p}\right)  ^{2}\mathbf{p}^{2}+8p_{0}%
^{2}\left(  \boldsymbol{\kappa}\cdot\mathbf{p}\right)  ^{2}.\label{ST2}%
\end{align}
Now, the full operator $\widetilde{D}^{\mu\nu}$ is written as:
\begin{equation}
\widetilde{D}^{\mu\nu}=-\left(
\begin{array}
[c]{ccccccc}%
p^{2} &  & p_{1}A-\kappa_{1}\mathbf{p}^{2} &  & p_{2}A-\kappa_{2}%
\mathbf{p}^{2} &  & p_{3}A-\kappa_{3}\mathbf{p}^{2}\\[0.2cm]%
p_{1}A-\kappa_{1}\mathbf{p}^{2} &  & -p^{2}-2p_{0}\left(  A-\kappa_{1}%
p_{1}\right)   &  & p_{0}\left(  \kappa_{1}p_{2}+\kappa_{2}p_{1}\right)   &  &
p_{0}\left(  \kappa_{3}p_{1}+\kappa_{1}p_{3}\right)  \\[0.2cm]%
p_{2}A-\kappa_{2}\mathbf{p}^{2} &  & p_{0}\left(  \kappa_{1}p_{2}+\kappa
_{2}p_{1}\right)   &  & -p^{2}-2p_{0}\left(  A-\kappa_{2}p_{2}\right)   &  &
p_{0}\left(  \kappa_{2}p_{3}+\kappa_{3}p_{2}\right)  \\[0.2cm]%
p_{3}A-\kappa_{3}\mathbf{p}^{2} &  & p_{0}\left(  \kappa_{3}p_{1}+\kappa
_{1}p_{3}\right)   &  & p_{0}\left(  \kappa_{2}p_{3}+\kappa_{3}p_{2}\right)
&  & -p^{2}-2p_{0}\left(  A-\kappa_{3}p_{3}\right)
\end{array}
\right)
\end{equation}
The propagator (\ref{propA}), in Fourier space, is given by the inverse of the
matrix $\widetilde{D}^{\mu\nu}$, which fulfills Eq. (\ref{PE3}). Being $M$ a
nonsingular matrix, its inverse is $M^{-1}=\left(  \det M\right)  ^{-1}%
$Adj$\left(  M\right)  =\left(  \det M\right)  ^{-1}\left(  \text{Cof}\left(
M\right)  \right)  ^{T}.$ Hence, the first step for the inversion procedure
consists in evaluating the determinant of matrix $\widetilde{D}^{\mu\nu}.$
Such determinant is composed of a sum of 76 different terms, however, such a
sum can be factorized as the product of three terms:
\begin{equation}
\det\widetilde{D}^{\mu\nu}=-p^{4}\left(  p^{2}+2Ap_{0}\right)  \left(
p^{2}+2Ap_{0}-\mathbf{p}^{2}\boldsymbol{\kappa}^{2}+A^{2}\right)
.\label{det2}%
\end{equation}

The propagator matrix, given as the inverse matrix of $\widetilde{D}^{\mu\nu
},$ is read at the form
\begin{equation}
\Delta_{\mu\nu}=\frac{1}{\boxplus}\left(
\begin{array}
[c]{ccccccc}%
-GH &  & GJ_{1} &  & GJ_{2} &  & GJ_{3}\\[0.2cm]%
GJ_{1} &  & C_{11} &  & M_{12} &  & M_{13}\\[0.2cm]%
GJ_{2} &  & M_{12} &  & C_{22} &  & M_{23}\\[0.2cm]%
GJ_{3} &  & M_{13} &  & M_{23} &  & C_{33}%
\end{array}
\right)  ,
\end{equation}
whose terms are defined as follows:%
\begin{align}
\boxplus &  =p^{4}\left(  p^{2}+2Ap_{0}\right)  \left(  p^{2}+2Ap_{0}%
-\mathbf{p}^{2}\boldsymbol{\kappa}^{2}+A^{2}\right)  ,\label{Term1a}\\[0.2cm]
\text{ }G &  =p^{2}+2p_{0}A,\text{ }F=\boldsymbol{\kappa}^{2}\mathbf{p}%
^{2}-A^{2},\\[0.2cm]
H &  =p^{4}+2p^{2}Ap_{0}-\left(  p_{0}\right)  ^{2}F,\\[0.2cm]
J_{i} &  =p^{2}\left[  \mathbf{p}^{2}\kappa_{i}-Ap_{i}\right]  +p_{0}%
p_{i}F,\label{Term1d}%
\end{align}%
\begin{align}
C_{ii}\;= &  \;p^{6}+2p^{4}\left[  A+\kappa_{i}p_{i}\right]  p_{0}%
+p^{2}\left[  4A\kappa_{i}p_{i}-\left(  \left[  \boldsymbol{\kappa}%
\times\mathbf{p}\right]  _{i}\right)  ^{2}\right]  \left(  p_{0}\right)
^{2}\nonumber\\[0.2cm]
&  +p^{2}\mathbf{p}^{2}F-p^{2}\left[  \mathbf{p}^{2}\kappa_{i}-Ap_{i}\right]
^{2}+2\left(  p_{i}\right)  ^{2}p_{0}A\left(  \boldsymbol{\kappa}%
\times\mathbf{p}\right)  ^{2},\label{Term2}%
\end{align}%
\begin{align}
M_{ij}\;=\; &  p^{4}\left(  \kappa_{i}p_{j}+\kappa_{j}p_{i}\right)
p_{0}-p^{2} \left[  \mathbf{p}^{2}\kappa_{i}-Ap_{i}\right]  \left[
\mathbf{p}^{2}\kappa_{j}-Ap_{j}\right] \nonumber\\[0.2cm]
&  +p^{2}\left[  2A\left(  \kappa_{i}p_{j}+\kappa_{j}p_{i}\right)  -\left(
\boldsymbol{\kappa}\times\mathbf{p}\right)  _{i}\left(  \boldsymbol{\kappa
}\times\mathbf{p}\right)  _{j}\right]  \left(  p_{0}\right)  ^{2}+2p_{i}%
p_{j}p_{0}A\left(  \boldsymbol{\kappa}\times\mathbf{p}\right)  ^{2}%
.\label{Term3}%
\end{align}
The relations\ (\ref{Term1a}-\ref{Term3}) reveal that the exact propagator is
a cumbersome algebraic expression. The manipulation of this full propagator,
however, yields (as an advantage) more reliable results. Indeed, it should be
mentioned that a first order propagator was initially carried out, but the
corresponding unitarity analysis has demonstrated to be inconsistent.

\subsubsection{Dispersion relations, causality and energy stability}

The dispersion relations for this electrodynamics are obtained from the poles
of the propagator, which in the case are given by $\boxplus=0.$ The factor
$\boxplus$ contains all information about the pole structure of the theory.
These relations are important to analyze the stability and causality of the
theory. Regarding the relation $\boxplus=0$, the dispersion relations are:
\begin{align}
\left[  p^{2}+2p_{0}\left(  \boldsymbol{\kappa}\cdot\mathbf{p}\right)
\right]   &  =0,\label{DR1}\\[0.2cm]
\left[  p^{2}+2p_{0}\left(  \boldsymbol{\kappa}\cdot\mathbf{p}\right)
-\boldsymbol{\kappa}^{2}\mathbf{p}^{2}+\left(  \boldsymbol{\kappa}%
\cdot\mathbf{p}\right)  ^{2}\right]   &  =0.\label{DR2}%
\end{align}
We first analyze relation (\ref{DR1}),
\begin{equation}
p_{0}^{2}-\mathbf{p}^{2}+2p_{0}\left(  \boldsymbol{\kappa}\cdot\mathbf{p}%
\right)  =0,\label{DR3}%
\end{equation}
whose roots are%
\begin{equation}
p_{0\pm}=-\left(  \boldsymbol{\kappa}\cdot\mathbf{p}\right)  \pm
\sqrt{\mathbf{p}^{2}+\left(  \boldsymbol{\kappa}\cdot\mathbf{p}\right)  ^{2}%
}.\label{E01}%
\end{equation}
Assuming that $|\mathbf{\kappa|<<}1,$ we have (at first order in
$|\mathbf{\kappa|)}$:%
\begin{align}
p_{0+} &  =|\mathbf{p|}-\left(  \boldsymbol{\kappa}\cdot\mathbf{p}\right)
,\label{E1A}\\[0.2cm]
p_{0-} &  =-(|\mathbf{p|}+\left(  \boldsymbol{\kappa}\cdot\mathbf{p}\right)
).\label{E1B}%
\end{align}
Here, the root $p_{0+}=|\mathbf{p|}-\left(  \boldsymbol{\kappa}\cdot
\mathbf{p}\right)  $ is positive definite since~$|\mathbf{\kappa|<<}1.$ On the
other hand, the mode $p_{0-}=-(|\mathbf{p|}+\left(  \boldsymbol{\kappa}%
\cdot\mathbf{p}\right)  )$ stands for an anti-particle, whose energy becomes
positive after reinterpretation, $p_{0-}=(|\mathbf{p|}+\left(
\boldsymbol{\kappa}\cdot\mathbf{p}\right)  ).$ In this sense, we can state
that the modes (\ref{E1A},\ref{E1B}) present positive energy, being stable.

The second dispersion relation is%
\begin{equation}
p_{0}^{2}-\mathbf{p}^{2}+2p_{0}\left(  \boldsymbol{\kappa}\cdot\mathbf{p}%
\right)  -\boldsymbol{\kappa}^{2}\mathbf{p}^{2}+\left(  \boldsymbol{\kappa
}\cdot\mathbf{p}\right)  ^{2}=0,\label{DR4B}%
\end{equation}
whose roots are the following:%
\begin{equation}
p_{0\pm}=-\left(  \boldsymbol{\kappa}\cdot\mathbf{p}\right)  \pm
\sqrt{\mathbf{p}^{2}(1+\boldsymbol{\kappa}^{2})}.\label{E02}%
\end{equation}

At first order in $|\mathbf{\kappa|,}$ we have attained the same expressions
(\ref{E1A},\ref{E1B}), which can be written simply as
\begin{equation}
p_{0\pm}=|\mathbf{p|}\mp\left(  \boldsymbol{\kappa}\cdot\mathbf{p}\right)
,\label{EFO}%
\end{equation}
We then conclude that this theory is endowed with energy stability (for small
values of $|\boldsymbol{\kappa}\mathbf{|).}$ It is important to mention that,
despite the double sign in the dispersion relations (\ref{EFO}), they yield
the same phase velocities for waves traveling at the same direction. Note the
the positive and negative frequency modes are associated with waves which
propagate in opposite directions and the term $\left(  \boldsymbol{\kappa
}\cdot\mathbf{p}\right)  $ changes the signal under the direction inversion
($\mathbf{p\rightarrow-p).}$ This result confirms the nonbirefringent
character of the coefficient $\kappa$ at leading order, as properly stated in
Refs. \cite{KM1,KM2,KM3,Kostelec}, and others \cite{Manojr1,Kob}.

Another issue of importance for consistency is the causality. At the quantum
level, causality is a feature that requires commutation between observables
separated by a spacelike interval (called microcausality in field theory). The
causality analysis, at a classical level, is related to the sign of the
propagator poles, given in terms of $\ p^{2},$ in such a way one must have
$p^{2}\geq0$ in order to preserve the causality (preventing the existence of
tachyons). Both dispersion relations (\ref{DR1}, \ref{DR2}) yield $p^{2}<0,$
which implies non-causal modes. A more detailed and confident analysis on
causality comes from the group velocity, $u_{g}=dp_{0}/d|\mathbf{p|},$ and
from the front velocity, $u_{front}=\lim_{|\mathbf{p|\rightarrow\infty}%
}u_{phase}$ (see Ref. \cite{Sexl}). The causality is assured if $u_{g}\leq1$
and $u_{front}\leq1.$ For relation (\ref{DR1}), we obtain%

\begin{equation}
u_{g}=\frac{dp_{0}}{d|\mathbf{p|}}=\frac{|\mathbf{p|}-p_{0}|\boldsymbol{\kappa
}\mathbf{\ ||p|}\cos\theta}{p_{0}+\left\vert \boldsymbol{\kappa}\right\vert
\left\vert \mathbf{p}\right\vert \cos\theta},
\end{equation}
and $u_{front}=(1\pm\left\vert \boldsymbol{\kappa}\right\vert \cos\theta),$
for $\boldsymbol{\kappa}\cdot\mathbf{p=}\left\vert \boldsymbol{\kappa
}\right\vert \left\vert \mathbf{p}\right\vert \cos\theta.$ Even for a small
background ($|\mathbf{\kappa|<<}1)$, it may occur that $|u_{g}\mathbf{|>}1$
and $u_{front}>1$ for some values of $p_{0},|\mathbf{p}|$. This is enough for
yielding causality violation. For the relation (\ref{DR2}), we have
\begin{equation}
u_{g}=\frac{dp_{0}}{d|\mathbf{p|}}=\frac{|\mathbf{p}|(1-\boldsymbol{\kappa
}^{2}\sin^{2}\theta)-2p_{0}|\boldsymbol{\kappa}\mathbf{|}\cos\theta}%
{p_{0}+|\boldsymbol{\kappa}\mathbf{||p|}\cos\theta},
\end{equation}
and $u_{front}=(1\pm|\boldsymbol{\kappa}|\cos\theta).$ At the same way, this
expression provides $|u_{g}\mathbf{|>}1,u_{front}>1$ for some values of
$\ p_{0},|\mathbf{p|,}$ which implies causality violation. Thus, we conclude
that this theory is stable but non-causal.

It is instructive to mention that the dispersion relations (\ref{E01},
\ref{E02}) can be obtained equivalently from the Maxwell equations, written
for this parametrization, as it appears in Ref. \cite{Manojr1}. Writing the
electric and magnetic fields in a Fourier representation, $\mathbf{B}%
(\mathbf{r})=\left(  2\pi\right)  ^{-3}\int\widetilde{\mathbf{B}}%
(\mathbf{p})\exp(-i\mathbf{p}\cdot\mathbf{r})d^{3}\mathbf{p}$, $\mathbf{E}%
(\mathbf{r})=\left(  2\pi\right)  ^{-3}\int\widetilde{\mathbf{E}}%
(\mathbf{p})\exp(-i\mathbf{p}\cdot\mathbf{r})d^{3}\mathbf{p}$, the Maxwell
equations take on the following form (at the absence of sources):
\begin{align}
\mathbf{p}\cdot\widetilde{\mathbf{E}} &  =-\mathbf{\boldsymbol{\kappa}}%
\cdot\left(  \mathbf{p}\times\widetilde{\mathbf{B}}\right)  ,\label{M1}%
\\[0.2cm]
\mathbf{p}\times\widetilde{\mathbf{B}}+p_{0}\left(  \widetilde{\mathbf{B}%
}\times\mathbf{\boldsymbol{\kappa}}\right)  +p_{0}\widetilde{\mathbf{E}} &
=-\mathbf{p}\times\left(  \widetilde{\mathbf{E}}\times
\mathbf{\boldsymbol{\kappa}}\right)  \mathbf{,}~\label{M2}\\[0.2cm]
\mathbf{p}\times\widetilde{\mathbf{E}}-p_{0}\widetilde{\mathbf{B}} &
=0,\text{ }\mathbf{p}\cdot\widetilde{\mathbf{B}}=0.\label{M3}%
\end{align}

From these expressions, it is attained an equation for the electric field
components, $\mathbb{M}^{jl}\widetilde{E}^{l}=0,$ where%
\begin{equation}
\mathbb{M}^{jl}=[p^{l}p^{j}-p_{0}p^{j}\kappa^{l}-p_{0}p^{l}\kappa^{j}%
+\delta^{jl}(p^{2}+2p_{0}p^{i}\kappa^{i})].
\end{equation}
Such operator can be represented as $3\times3$ matrix,
\begin{equation}
\mathbb{M}=\left[
\begin{array}
[c]{ccccccc}%
p^{2}+2p_{0}A+p_{1}^{2}-2p_{0}p_{1}\kappa_{1} &  &  & p_{1}p_{2}-p_{0}%
p_{1}\kappa_{2}-p_{0}p_{2}\kappa_{1} &  &  & p_{1}p_{3}-p_{0}p_{1}\kappa
_{3}-p_{0}p_{3}\kappa_{1}\\[0.25cm]%
p_{1}p_{2}-p_{0}p_{1}\kappa_{2}-p_{0}p_{2}\kappa_{1} &  &  & p^{2}%
+2p_{0}A+p_{2}^{2}-2p_{0}p_{2}\kappa_{2} &  &  & p_{2}p_{3}-p_{0}p_{2}%
\kappa_{3}-p_{0}p_{3}\kappa_{2}\\[0.25cm]%
p_{1}p_{3}-p_{0}p_{1}\kappa_{3}-p_{0}p_{3}\kappa_{1} &  &  & p_{2}p_{3}%
-p_{0}p_{2}\kappa_{3}-p_{0}p_{3}\kappa_{2} &  &  & p^{2}+2p_{0}A+p_{3}%
^{2}+2p_{0}p_{3}\kappa_{3}%
\end{array}
\right]  ,
\end{equation}
whose determinant is a sum of 60 terms. After suitable simplification, this
determinant takes the form%
\begin{equation}
\det\mathbb{M}=p_{0}^{2}\left(  p^{2}+2Ap_{0}\right)  \left(  p^{2}%
+2Ap_{0}-\mathbf{p}^{2}\boldsymbol{\kappa}^{2}+A^{2}\right)  .
\end{equation}

The condition $\det\mathbb{M}=0\ $provides the non-trivial solutions for Eq.
(\ref{M3}) and the associated dispersion relations of this model. This
alternative procedure confirms the correctness of dispersion relations
(\ref{DR1}, \ref{DR2}), in which the present consistency analysis is based.

\subsubsection{The unitarity analysis}

The unitarity analysis of this model at tree-level is here carried out through
the saturation of the propagators with external currents, which must be
implemented by means of the saturated propagator ($SP$), a scalar quantity
given as follows:
\begin{equation}
SP=J^{\ast\mu}\text{Res}(\Delta_{\mu\nu})\text{ }J^{\nu},\label{Sat2}%
\end{equation}
where Res$(\Delta_{\mu\nu})$ is the matrix residue evaluated at the pole of
the propagator. The gauge current $(J^{\mu})$\ satisfies the conservation law
$\left(  \partial_{\mu}J^{\mu}=0\right)  ,$\ which in momentum space is read
as $p_{\mu}J^{\mu}=0$. In accordance with this method, the unitarity analysis
is assured whenever the imaginary part of the saturation $SP$\ (at the poles
of the propagator) is positive (for further details see Ref. \cite{Veltman}).
This method was applied in some Lorentz-violating models \cite{Belich},
\cite{Unitarity}. A way to carry out the saturation consists in determining
the eigenvalues of the propagator matrix, evaluated at its own poles.

We begin analyzing the unitarity for the pole associated with Eq. (\ref{DR1}),
for which $p^{2}=-2p_{0}\left(  \boldsymbol{\kappa}\cdot\mathbf{p}\right)  $.
Without loss of generality, we adopt the four-momentum $p_{\mu}=(p_{0}%
,0,0,p_{3})$, for which we have $A=\kappa_{3}p_{3},$ and $p_{3}^{2}=p_{0}%
^{2}+2p_{0}p_{3}\kappa_{3}$, $F=(\kappa_{1}^{2}+\kappa_{2}^{2})p_{3}^{2}.$ At
this pole the propagator matrix is written as:
\begin{equation}
\Delta_{\mu\nu}=R_{2}\left(
\begin{array}
[c]{ccccccc}%
0 &  & 0 &  & 0 &  & 0\\[0.2cm]%
0 &  & C_{11} &  & M_{12} &  & 0\\[0.2cm]%
0 &  & M_{12} &  & C_{22} &  & 0\\[0.2cm]%
0 &  & 0 &  & 0 &  & 0
\end{array}
\right)  ,\label{R1}%
\end{equation}
where $R_{2}=-[4p_{0}^{2}\kappa_{3}^{4}(\kappa_{1}^{2}+\kappa_{2}^{2}%
)p_{3}^{2}]^{-1}$ is the residue of $\boxplus^{-1}$carried out at the pole
$p^{2}=-2p_{0}\left(  p_{3}\kappa_{3}\right)  ,$ and
\begin{equation}
C_{11}=-4p_{0}^{2}p_{3}^{4}\kappa_{3}^{2}\kappa_{2}^{2},\text{ }C_{22}%
=-4p_{0}^{2}p_{3}^{4}\kappa_{3}^{2}\kappa_{1}^{2},\text{ }M_{12}=4\kappa
_{1}\kappa_{2}\kappa_{3}^{2}p_{3}^{4}p_{0}^{2}.
\end{equation}
The eigenvalues of matrix (\ref{R1}) are $\lambda_{1}=\lambda_{2}=\lambda
_{3}=0,$ $\lambda_{4}=-4p_{0}^{2}p_{3}^{4}\kappa_{3}^{2}(\kappa_{2}^{2}%
+\kappa_{1}^{2}).$ We have attained a negative eigenvalue and a negative
residue $R_{1}$. Hence, the saturation turns out positive, implying unitarity
preservation at this pole for any $\mathbf{\kappa=}(\kappa_{1},\kappa
_{2},\kappa_{3})$.

A similar analysis can be performed for the pole associated with Eq.
(\ref{DR2}), $p^{2}+2p_{0}\left(  \boldsymbol{\kappa}\cdot\mathbf{p}\right)
=\boldsymbol{\kappa}^{2}\mathbf{p}^{2}-\left(  \boldsymbol{\kappa}%
\cdot\mathbf{p}\right)  ^{2}$. Following the prescription $p_{\mu}%
=(p_{0},0,0,p_{3}),$ we have $p_{0}^{2}=p_{3}^{2}(1+\kappa_{\perp}%
^{2})+2\allowbreak p_{0}p_{3}\kappa_{3}.$ The propagator matrix, at this pole,
is read as
\begin{equation}
\Delta_{\mu\nu}=R_{3}\left(
\begin{array}
[c]{cccccccccc}%
\kappa_{\perp}^{4}p_{3}^{6} &  &  & \kappa_{\perp}^{2}p_{3}^{4}\kappa_{1}p^{2}
&  &  & \kappa_{\perp}^{2}p_{3}^{4}\kappa_{2}p^{2} &  &  & p_{0}^{2}%
\kappa_{\perp}^{4}p_{3}^{5}\\[0.3cm]%
\kappa_{\perp}^{2}p_{3}^{4}\kappa_{1}p^{2} &  &  & p^{4}\kappa_{1}^{2}%
p_{3}^{2} &  &  & p^{4}p_{3}^{2}\kappa_{1}\kappa_{2} &  &  & p^{2}p_{0}%
p_{3}^{3}\kappa_{\perp}^{2}\kappa_{1}\\[0.3cm]%
\kappa_{\perp}^{2}p_{3}^{4}\kappa_{2}p^{2} &  &  & p^{4}p_{3}^{2}\kappa
_{1}\kappa_{2} &  &  & p^{4}\kappa_{2}^{2}p_{3}^{2} &  &  & p^{2}p_{0}%
p_{3}^{3}[\kappa_{\perp}^{2}\kappa_{2}]\\[0.3cm]%
p_{0}\kappa_{\perp}^{4}p_{3}^{5} &  &  & p^{2}p_{0}p_{3}^{3}\kappa_{\perp}%
^{2}\kappa_{1} &  &  & p^{2}p_{0}p_{3}^{3}[\kappa_{\perp}^{2}\kappa_{2}] &  &
& C_{33}%
\end{array}
\right)  ,\label{R2}%
\end{equation}
where $\kappa_{\perp}^{2}=\kappa_{1}^{2}+\kappa_{2}^{2},C_{33}=p^{2}p_{3}%
^{4}[\kappa_{\perp}^{4}+\kappa_{\perp}^{2}]+2p_{0}\kappa_{3}p_{3}^{5}%
\kappa_{\perp}^{2},$ and $R_{3}=[(2p_{0}p_{3}\kappa_{3}-\kappa_{\perp}%
^{2}p_{3}^{2})^{2}\kappa_{\perp}^{2}p_{3}^{2}]^{-1}$ is the residue of
$\boxplus^{-1}$ carried out at the pole $p^{2}=-2p_{0}\left(
\boldsymbol{\kappa}\cdot\mathbf{p}\right)  +\boldsymbol{\kappa}^{2}%
\mathbf{p}^{2}-\left(  \boldsymbol{\kappa}\cdot\mathbf{p}\right)  ^{2}.$ Given
the structure of this matrix, a simpler analysis is first performed for
$\mathbf{\kappa=}\left(  \kappa_{1},0,0\right)  .$ In this case, the matrix
(\ref{R2}) takes the form:
\begin{equation}
\Delta_{\mu\nu}=R_{3}\left(
\begin{array}
[c]{ccccccc}%
\kappa_{1}^{4}p_{3}^{6} &  & \kappa_{1}^{3}p_{3}^{4}p^{2} &  & 0 &  &
p_{0}\kappa_{1}^{4}p_{3}^{5}\\[0.25cm]%
\kappa_{1}^{3}p_{3}^{4}p^{2} &  & p_{3}^{2}p^{4}\kappa_{1}^{2} &  & 0 &  &
p^{2}p_{0}p_{3}^{3}\kappa_{1}^{3}\\[0.25cm]%
0 &  & 0 &  & 0 &  & 0\\[0.25cm]%
p_{0}\kappa_{1}^{4}p_{3}^{5} &  & p^{2}p_{0}p_{3}^{3}\kappa_{1}^{3} &  & 0 &
& p_{3}^{6}(\kappa_{1}^{6}+\kappa_{1}^{4})
\end{array}
\right)  ,\label{R3}%
\end{equation}
which has a unique non-null eigenvalue: $\lambda=2p_{3}^{6}\kappa_{1}%
^{4}(\kappa_{1}^{2}+1).$ As this eigenvalue and $R_{3}$ are both positive, the
unitarity is assured for this particular case. The same inspection can be
performed for the case $\mathbf{\kappa}=\left(  0,\kappa_{2},0\right)  ,$ for
which the resulting matrix has also only a non-null eigenvalue: $\lambda=$
$2p_{3}^{6}\kappa_{2}^{4}(\kappa_{2}^{2}+1).$ For the particular case,
$\kappa=\left(  0,0,\kappa_{3}\right)  ,$ the matrix is reduced to a null
matrix, which is consistent with unitarity preservation.\ By the results
attained for the particular configurations $\left(  \kappa_{1},0,0\right)  ,$
$\left(  0,\kappa_{2},0\right)  ,$ $\left(  0,0,\kappa_{3}\right)  ,$ we infer
that the unitarity holds for the general case\ $\left(  \kappa_{1},\kappa
_{2},\kappa_{3}\right)  $ as well. Hence, we conclude that the excitations
stemming from this pole are unitary. For an alternative analysis of the
unitarity see the Appendix \ref{apendice2}.

Finally, we conclude that the physical modes of the parity-odd sector,
represented by the coefficients $\kappa^{j}$, imply a unitary electrodynamics.

\subsection{The parity-even gauge propagator}

We now consider the evaluation of the propagator for the electrodynamics
associated with the parity-even sector of the tensor $W_{\alpha\nu\rho\varphi
}.$ In this case, we are interested in the nonbirefringent components of the
parity-even sector, represented by the elements of the matrix $\widetilde
{\kappa}_{e-}$ and the trace element. The classical solutions for the Maxwell
electrodynamics supplemented by this term were recently analyzed in Ref.
\cite{Paulo}. In order to isolate the parity-even sector, we take as null the
parity-odd sector $\left(  \kappa_{DB}=\kappa_{HE}=0\right)  $. The
parity-even sector is constrained by the condition $\left(  \kappa
_{DE}=-\kappa_{HB}\right)  ,$ which implies $\widetilde{\kappa}_{e+}=0$
(eliminating the birefringent components). The remaining elements
(nonbirefringent ones) are located in matrix $\left(  \widetilde{\kappa}%
_{e-}\right)  ,$ whose components obey the following parametrization:
\begin{equation}
(\kappa_{DE})^{jk}=\left(  \widetilde{\kappa}_{e-}\right)  ^{jk}%
+\kappa_{\text{tr}}\delta^{jk},\text{ }%
\end{equation}
where $\kappa_{\text{tr}}=\frac{1}{3}$tr$(\kappa_{DE})$ is the trace element.
Under this parametrization, the nonnull elements of the tensor $W^{\mu
\alpha\beta\nu}$ are
\begin{equation}
W_{0j0k}=-\frac{1}{2}\left(  \widetilde{\kappa}_{e-}\right)  _{ij}%
-\frac{\kappa_{\text{tr}}}{2}\delta_{ij},
\end{equation}%
\begin{equation}
W_{pqlm}=-\frac{1}{2}\epsilon_{pqj}\epsilon_{lmk}\left(  \widetilde{\kappa
}_{e-}\right)  _{jk}-\frac{\kappa_{\text{tr}}}{2}\left[  \delta_{pl}%
\delta_{qm}-\delta_{pm}\delta_{ql}\right]  .
\end{equation}

For $i\neq j$ the elements $\left(  \kappa_{DE}\right)  ^{ij}$ imply vacuum
anisotropy while $\kappa_{\text{tr}}$ is compatible with an isotropic and
homogeneous space. For preserving the space isotropy, one should retain only
the trace element, $\kappa_{\text{tr}},$ taking as null the non-diagonal
elements [$\left(  \kappa_{DE}\right)  ^{12}=\left(  \kappa_{DE}\right)
^{13}=\left(  \kappa_{DE}\right)  ^{23}=0]$. This model is then represented by
the following Lagrangian:%
\begin{equation}
\mathcal{L}=\frac{1}{2}\left[  \left(  1+\kappa_{\text{tr}}\right)
\mathbf{E}^{2}-\left(  1-\kappa_{\text{tr}}\right)  \mathbf{B}^{2}\right]
-\rho A_{0}+\mathbf{j\cdot A}.\label{Ln1}%
\end{equation}
Thus, the parameter $\kappa_{\text{tr}}$ is known as the isotropic LIV
parameter. An investigation involving this unique term was developed in Ref.
\cite{Klink3}, where it was set up a two-sided bound on it from data
confirming the absence of Cherenkov radiation emitted from UHECRs. In this
case, the operator (\ref{S01}) is written as
\begin{equation}
\tilde{S}^{\mu\nu}=\left(
\begin{array}
[c]{ccccccc}%
\kappa_{\text{tr}}(p_{1}^{2}+p_{2}^{2}+p_{3}^{2}) &  & -\kappa_{\text{tr}%
}p_{1}p_{0} &  & -\kappa_{\text{tr}}p_{2}p_{0} &  & -\kappa_{\text{tr}}%
p_{3}p_{0}\\[0.2cm]%
-\kappa_{\text{tr}}p_{1}p_{0} &  & \kappa_{\text{tr}}\left(  p_{0}^{2}%
+p_{2}^{2}+p_{3}^{2}\right)   &  & -\kappa_{\text{tr}}p_{1}p_{2} &  &
-\kappa_{\text{tr}}p_{3}p_{1}\\[0.2cm]%
-\kappa_{\text{tr}}p_{2}p_{0} &  & -\kappa_{\text{tr}}p_{1}p_{2} &  &
\kappa_{\text{tr}}\left(  p_{0}^{2}+p_{1}^{2}+p_{3}^{2}\right)   &  &
-\kappa_{\text{tr}}p_{3}p_{2}\\[0.2cm]%
-\kappa_{\text{tr}}p_{3}p_{0} &  & -\kappa_{\text{tr}}p_{3}p_{1} &  &
-\kappa_{\text{tr}}p_{3}p_{2} &  & \kappa_{\text{tr}}\left(  p_{0}^{2}%
+p_{1}^{2}+p_{2}^{2}\right)
\end{array}
\right)  ,
\end{equation}
whereas the full matrix operator $\widetilde{D}^{\mu\nu}$ is
\begin{equation}
\widetilde{D}^{\mu\nu}=-\left(
\begin{array}
[c]{ccccccc}%
p^{2}-\kappa_{\text{tr}}(p_{1}^{2}+p_{2}^{2}+p_{3}^{2}) &  & \kappa
_{\text{tr}}p_{1}p_{0} &  & \kappa_{\text{tr}}p_{2}p_{0} &  & \kappa
_{\text{tr}}p_{3}p_{0}\\[0.2cm]%
\kappa_{\text{tr}}p_{1}p_{0} &  & -p^{2}-\kappa_{\text{tr}}\left(  p_{0}%
^{2}+p_{2}^{2}+p_{3}^{2}\right)   &  & \kappa_{\text{tr}}p_{1}p_{2} &  &
\kappa_{\text{tr}}p_{3}p_{1}\\[0.2cm]%
\kappa_{\text{tr}}p_{2}p_{0} &  & \kappa_{\text{tr}}p_{1}p_{2} &  &
-p^{2}-\kappa_{\text{tr}}\left(  p_{0}^{2}+p_{1}^{2}+p_{3}^{2}\right)   &  &
\kappa_{\text{tr}}p_{3}p_{2}\\[0.2cm]%
\kappa_{\text{tr}}p_{3}p_{0} &  & \kappa_{\text{tr}}p_{3}p_{1} &  &
\kappa_{\text{tr}}p_{3}p_{2} &  & -p^{2}-\kappa_{\text{tr}}\left(  p_{0}%
^{2}+p_{1}^{2}+p_{2}^{2}\right)
\end{array}
\right)  .
\end{equation}

The determinant of this matrix is a sum of 35 different terms which can be
carefully simplified to the form:
\begin{equation}
\det\widetilde{D}^{\mu\nu}=-\left(  1+\kappa_{\text{tr}}\right)  p^{4}\left[
\left(  \kappa_{\text{tr}}+1\right)  p^{2}+2\kappa_{\text{tr}}\mathbf{p}%
^{2}\right]  ^{2}.
\end{equation}
The propagator matrix can be exactly evaluated in terms of $\kappa_{\text{tr}%
}$, being read at the form%
\begin{equation}
\Delta_{\mu\nu}=-\frac{1}{\boxdot}\left(
\begin{array}
[c]{ccccccc}%
(\left(  \kappa_{\text{tr}}+1\right)  p^{2}+\kappa_{\text{tr}}\mathbf{p}%
^{2})M &  & \kappa_{\text{tr}}p_{0}p_{1}M &  & \kappa_{\text{tr}}p_{0}p_{2}M &
& \kappa_{\text{tr}}p_{0}p_{3}M\\[0.2cm]%
\kappa_{\text{tr}}p_{0}p_{1}M &  & -P_{11} &  & -\kappa_{\text{tr}}p_{1}%
p_{2}C &  & -\kappa_{\text{tr}}p_{1}p_{3}C\\[0.2cm]%
\kappa_{\text{tr}}p_{0}p_{2}M &  & -\kappa_{\text{tr}}p_{1}p_{2}C &  & -P_{22}
&  & -\kappa_{\text{tr}}p_{2}p_{3}C\\[0.2cm]%
\kappa_{\text{tr}}p_{0}p_{3}M &  & -\kappa_{\text{tr}}p_{1}p_{3}C &  &
-\kappa_{\text{tr}}p_{2}p_{3}C &  & -P_{33}%
\end{array}
\right)  ,
\end{equation}
where%
\begin{align}
M &  =\left[  \left(  1+\kappa_{\text{tr}}\right)  p^{2}+2\kappa_{\text{tr}%
}\mathbf{p}^{2}\right]  ,\text{ }P_{ii}=[\left(  1+\kappa_{\text{tr}}\right)
p^{4}+\kappa_{\text{tr}}\left(  p_{i}\right)  ^{2}C],\\
C &  =\left[  \left(  1-\kappa_{\text{tr}}\right)  p^{2}-2\kappa_{\text{tr}%
}\mathbf{p}^{2}\right]  ,\text{ }\boxdot=\left(  1+\kappa_{\text{tr}}\right)
p^{4}M.\label{delta}%
\end{align}

\subsubsection{Dispersion relations and consistency analysis}

$\allowbreak$The dispersion relations for this model are obtained from the
poles of the propagator, read off from $\boxdot=0.$ These relations are
important to analyze the energy positivity (stability) and causality of the
theory. Considering the expression (\ref{delta}), the dispersion relations
are
\begin{equation}
(1+\kappa_{\text{tr}})p_{0}^{2}-(1-\kappa_{\text{tr}})\mathbf{p}^{2}=0,
\end{equation}
whose roots are
\begin{equation}
p_{0}=\pm\sqrt{\frac{1-\kappa_{\text{tr}}}{1+\kappa_{\text{tr}}}}%
|\mathbf{p|}.\label{DR4}%
\end{equation}
The relation (\ref{DR4}) reveals a model without birefringence, once the
positive and negative frequency modes propagates with the same phase velocity
$\left(  u_{ph}=\sqrt{(1-\kappa_{\text{tr}})/(1+\kappa_{\text{tr}})}\right)
.$ This result is in accordance with Refs. \cite{KM1,KM2,KM3,Kostelec}. It
also shows that the light velocity is less than 1. The energy stability of
this model is assured, once the energy of both modes is positive (after reinterpretation).

The causality of these modes seems to be spoiled, since relation~(\ref{DR4})
provides%
\begin{equation}
p^{2}=-\frac{2\kappa_{\text{tr}}}{(1-\kappa_{\text{tr}})}p_{0}^{2}<0.
\end{equation}
As this criterion is not enough to spoil causality, the group velocity
($u_{g}=dp_{0}/d|\mathbf{p|})$ shall be evaluated:
\begin{equation}
u_{g}=\pm\sqrt{\frac{1-\kappa_{\text{tr}}}{1+\kappa_{\text{tr}}}},
\end{equation}
It is less then 1 $\left(  u_{g}<1\right)  $ for $0<\kappa_{\text{tr}}<1.$
This theory presents equal phase, group and front velocities, as a consequence
of the non-dispersive relation (\ref{DR4})\textbf{. }We thus conclude that
this model has stability and causality assured for $0\leq\kappa_{\text{tr}%
}<1.$

As for the unitarity issue, we consider now the first order pole $[\left(
\kappa_{\text{tr}}+1\right)  p^{2}+2\kappa_{\text{tr}}\mathbf{p}^{2}]=0$,
whose residue at $\boxdot^{-1}$is $-1/(16\kappa_{\text{tr}}\mathbf{p}^{4}). $
At this pole, it holds $M=0$, $p^{2}=-2\kappa_{\text{tr}}\mathbf{p}%
^{2}/\left(  \kappa_{\text{tr}}+1\right)  ,C=-4\kappa_{\text{tr}}%
\mathbf{p}^{2}/\left(  \kappa_{\text{tr}}+1\right)  .$ For $p_{\mu}%
=(p_{0},0,0,p_{3}),$ the residue matrix is read as
\begin{equation}
\Delta_{\mu\nu}=\frac{1}{\left(  1+\kappa_{\text{tr}}\right)  }\left(
\begin{array}
[c]{cccc}%
0 & 0 & 0 & 0\\
0 & 1 & 0 & 0\\
0 & 0 & 1 & 0\\
0 & 0 & 0 & 0
\end{array}
\right)  .
\end{equation}
As the eigenvalues are $0,1,1,0$, the saturation $SP$ turns out positive and
this pole preserves unitarity for $\kappa_{\text{tr}}$ bounded in the range
$\kappa_{\text{tr}}>-1$.\ Moreover, the coefficient $\kappa_{\text{tr}}$ is
bounded in the range $0<\kappa_{\text{tr}}<1$ for assuring both causality and
unitarity. These results show that the physical excitations of this model are
defined for $0\leq\kappa_{\text{tr}}<1$ and are the ones associated with the
dispersion relation (\ref{DR4}). Thus, we have verified that the physical
modes of the theory represented by Lagrangian (\ref{Ln1}) are stable, causal
and unitary only for values of $\kappa_{\text{tr}}$ in the range $0\leq
\kappa_{\text{tr}}<1.$ In Appendix \ref{apendice2}, we give an alternative
analysis for the unitarity.

\section{ Conclusions}

In this work, we have evaluated the gauge propagator for the CPT-even sector
of the standard model extension. We have started carrying out the gauge
propagator for the parity-odd part of tensor $W_{\alpha\nu\rho\varphi},$
considering as non null only three components of this sector, according to the
parametrization stated in Refs. \cite{Manojr1,Kob}. This propagator was
carried out as a $4\times4$ matrix, whose poles were used to write the
dispersion relations and to investigate causality, energy stability and
unitarity. As a result, it was demonstrated that the electrodynamics
represented by these three coefficients is stable, non-causal and unitary.

The same procedure was applied to the parity-even part of tensor $W_{\alpha
\nu\rho\varphi}$. In this case, from the six nonbirefringent components, it
was retained only the isotropic ($\kappa_{\text{tr}}$) parameter. The
propagator was also written as a $4\times4$ matrix and the dispersion
relations were determined. The physical modes\textbf{\ }of this
electrodynamics revealed to be causal, stable and unitary for $0\leq
\kappa_{\text{tr}}<1$. It is important to point out that the causality,
according to the criterion of group and front velocities, is assured only for
positive values of $\kappa_{\text{tr}}.$

It is worthy to mention that this parity-even electrodynamics has been
recently investigated in a quantum electrodynamics (QED) environment, focusing
on the fermion-photon vertex interaction. This issue has connections with the
emission of Cherenkov radiation by rapid fermions (for the case $\kappa
_{\text{tr}}>0)$ or the photon decaying into a fermion-antifermion pair (for
the case $\kappa_{\text{tr}}<0)$ \cite{Klink3,Interac3}. The possibility of
having a negative $\kappa_{\text{tr}}$ in this context is opened by a
coordinate rescaling (see Appendix of Refs. \cite{Bailey,Interac3}) showing
that the nonbirefringent coefficients $k^{\mu\nu}=W_{\alpha}^{\text{ \ }%
\mu\alpha\nu}$ are physically equivalent to the electron sector coefficients
$c_{e}^{\mu\nu}$ in the context of a QED involving the isotropic CPT-even
electrodynamics and the fermion sector of the SME \cite{Fermion,Lehnert}.
Here, $\kappa_{\text{tr}}= $ $-2k^{00}/3.$ This states an equivalence between
$\kappa_{\text{tr}}\ $and $c_{e}^{00}$, in such a way that only the quantity
$\kappa_{\text{tr}}-4c_{e}^{00}/3$ is physically observable. Taking
$c_{e}^{00}=0,$ the possibility of having $\kappa_{\text{tr}}<0$ becomes
meaningful (see Refs. \cite{Klink3,Interac3}). However, the causality issue
for $\kappa_{\text{tr}}<0$ remains to be more discussed, once it clearly fails
in its strong version (which requires transmission of physical signs below the
light velocity). The authors of Ref. \cite{Interac3} support that causality
can be ensured working with a rescaled Minkowski spacetime in which photons
and fermions move in different light cones (see footnote [44] of Ref.
\cite{Interac3}).

Finally, the evaluation of the gauge propagator of this theory opens some
interesting possibilities of investigation. A feasible one seems to be the
coupling of this gauge theory with Dirac fermions, which allows the
investigation of fermion scattering and fermion decays processes mediated by
the LIV modified gauge sector. Such calculation may reveal how the LIV
coefficients affect some well known results of QED (quantum electrodynamics),
providing a new way to constrain the magnitude of the LIV parameters.

\appendix

\section{The double pole in $p^{2}=0$ \label{apendice1}}

As a matter of fact, we state that the excitations associated with the double
pole $\left(  p^{2}\right)  ^{2}=0$\ are nonphysical in both sectors
(parity-odd and parity-even). This conclusion comes primarily from its
dependence on the gauge fixing parameter $\left(  \xi\right)  $. The first
step is to evaluate the gauge propagator incorporating an arbitrary parameter
$\xi.$ We begin considering the parity-odd sector. The components of gauge
field propagator are%
\begin{equation}
\tilde{\Delta}_{00}=-\frac{1}{p^{2}}+\frac{\mathbf{p}^{2}\mathbf{R}}%
{p^{4}\left[  p^{2}+2p_{0}\left(  \mathbf{\kappa\cdot p}\right)
-\mathbf{R}\right]  }+\frac{\lambda}{\left(  1+\lambda\right)  p^{2}}%
+\frac{\lambda\mathbf{p}^{2}}{\left(  1+\lambda\right)  p^{4}},
\end{equation}%
\begin{align}
\tilde{\Delta}_{0j}  & =\frac{\mathbf{p}^{2}\kappa_{j}-\left(  \mathbf{\kappa
\cdot p}\right)  p_{j}}{p^{2}\left[  p^{2}+2p_{0}\left(  \mathbf{\kappa\cdot
p}\right)  -\mathbf{R}\right]  }+\frac{p_{0}p_{j}\mathbf{R}}{p^{4}\left[
p^{2}+2p_{0}\left(  \mathbf{\kappa\cdot p}\right)  -\mathbf{R}\right]  }\\
& +\lambda\frac{p_{0}+2\left(  \mathbf{\kappa\cdot p}\right)  }{\left(
1+\lambda\right)  p^{2}\left[  p^{2}+2p_{0}\left(  \mathbf{\kappa\cdot
p}\right)  -\mathbf{R}\right]  }p_{j}+\lambda\frac{2\left(  \mathbf{\kappa
\cdot p}\right)  \mathbf{p}^{2}-p_{0}\mathbf{R}}{\left(  1+\lambda\right)
p^{4}\left[  p^{2}+2p_{0}\left(  \mathbf{\kappa\cdot p}\right)  -\mathbf{R}%
\right]  }p_{j},\nonumber
\end{align}%
\begin{align}
\tilde{\Delta}_{ij}\left(  p\right)   & =\frac{\delta_{ij}}{p^{2}%
+2p_{0}\left(  \mathbf{\kappa\cdot p}\right)  }+~\frac{\mathbf{p}^{2}}{\left[
p^{2}+2p_{0}\left(  \mathbf{\kappa\cdot p}\right)  \right]  \left[
p^{2}+2p_{0}\left(  \mathbf{\kappa\cdot p}\right)  -\mathbf{R}\right]  }%
\kappa_{i}\kappa_{j}\nonumber\\
& +\left[  \frac{p_{0}}{p^{2}\left[  p^{2}+2p_{0}\left(  \mathbf{\kappa\cdot
p}\right)  -\mathbf{R}\right]  }-\frac{\left(  \mathbf{\kappa\cdot p}\right)
}{\left[  p^{2}+2p_{0}\left(  \mathbf{\kappa\cdot p}\right)  \right]  \left[
p^{2}+2p_{0}\left(  \mathbf{\kappa\cdot p}\right)  -\mathbf{R}\right]
}\right]  \left(  \kappa_{i}p_{j}+\kappa_{j}p_{i}\right) \nonumber\\
& +\left[  \frac{\mathbf{\kappa}^{2}\ }{\left[  p^{2}+2p_{0}\left(
\mathbf{\kappa\cdot p}\right)  \right]  \left[  p^{2}+2p_{0}\left(
\mathbf{\kappa\cdot p}\right)  -\mathbf{R}\right]  }+\frac{\mathbf{R}}%
{p^{4}\left[  p^{2}+2p_{0}\left(  \mathbf{\kappa\cdot p}\right)
-\mathbf{R}\right]  }+\frac{\lambda}{\left(  1+\lambda\right)  p^{4}}\right]
p_{i}p_{j}.
\end{align}
where $\lambda=(1/\xi-1)$ and $R=p^{2}\kappa^{2}-\left(  \mathbf{\kappa\cdot
p}\right)  ^{2}.$\ The residues in the double pole $p^{4}=0$\ are given by
\begin{align}
\text{Res}\left[  \tilde{\Delta}_{00}\right]   & =-\frac{1}{\left(
1+\lambda\right)  }-\frac{\mathbf{p}^{2}\mathbf{R}}{\left[  2p_{0}\left(
\mathbf{\kappa\cdot p}\right)  -\mathbf{R}\right]  ^{2}},\\
\text{Res}\left[  \tilde{\Delta}_{0j}\right]   & =\frac{\mathbf{p}^{2}%
\kappa_{j}-\left(  \mathbf{\kappa\cdot p}\right)  p_{j}}{\left[  2p_{0}\left(
\mathbf{\kappa\cdot p}\right)  -\mathbf{R}\right]  }-\frac{p_{0}%
p_{j}\mathbf{R}}{\left[  2p_{0}\left(  \mathbf{\kappa\cdot p}\right)
-\mathbf{R}\right]  ^{2}}+\lambda\frac{2\left(  \mathbf{\kappa\cdot p}\right)
}{\left(  1+\lambda\right)  \left[  2p_{0}\left(  \mathbf{\kappa\cdot
p}\right)  -\mathbf{R}\right]  }p_{j},\\
\text{Res}\left[  \tilde{\Delta}_{ij}\left(  p\right)  \right]   &
=\frac{p_{0}}{\left[  2p_{0}\left(  \mathbf{\kappa\cdot p}\right)
-\mathbf{R}\right]  }\left(  \kappa_{i}p_{j}+\kappa_{j}p_{i}\right)
-\frac{\mathbf{R}}{\left[  2p_{0}\left(  \mathbf{\kappa\cdot p}\right)
-\mathbf{R}\right]  ^{2}}p_{i}p_{j}%
\end{align}
By using the current conservation condition in momentum space, $p_{\mu}%
\tilde{J}^{\mu}=0$, and the fact $p^{2}=0$, the saturation reads as
\begin{equation}
SP=\left(  \tilde{J}^{0}\right)  ^{2}\left\{  \frac{\mathbf{R}-\lambda
2p_{0}\left(  \mathbf{\kappa\cdot p}\right)  }{\left(  1+\lambda\right)
\left[  2p_{0}\left(  \mathbf{\kappa\cdot p}\right)  -\mathbf{R}\right]
}\right\}  .
\end{equation}
As it is gauge-dependent, the pole $p^{2}=0$\ is nonphysical.

In the parity-even case, the components of the gauge field propagator are
given by%
\begin{align}
\tilde{\Delta}_{00}  & =-i\frac{\left(  \kappa_{\text{tr}}+1\right)
p^{2}+\left(  \kappa_{\text{tr}}-\lambda\right)  \mathbf{p}^{2}}{\left(
1+\lambda\right)  \left(  \kappa_{\text{tr}}+1\right)  p^{4}}%
~\ \ \ ,~\ \ \ \tilde{\Delta}_{0k}=\tilde{\Delta}_{0k}=-i\frac{\left(
\kappa_{\text{tr}}-\lambda\right)  p_{0}p_{k}}{\left(  \kappa_{\text{tr}%
}+1\right)  \left(  1+\lambda\right)  p^{4}},\\
\tilde{\Delta}_{jk}  & =i\frac{\delta_{jk}}{\left[  \left(  \kappa_{\text{tr}%
}+1\right)  p^{2}+2\kappa_{\text{tr}}\mathbf{p}^{2}\right]  }+i\frac
{p_{j}p_{k}\left\{  \kappa_{\text{tr}}\left[  \left(  1-\kappa_{\text{tr}%
}\right)  p^{2}-2\kappa_{\text{tr}}\mathbf{p}^{2}\right]  +\lambda\left[
\left(  1+3\kappa_{\text{tr}}\right)  p^{2}+2\kappa_{\text{tr}}\mathbf{p}%
^{2}\right]  \right\}  }{\left(  \kappa_{\text{tr}}+1\right)  \left(
1+\lambda\right)  p^{4}\left[  \left(  \kappa_{\text{tr}}+1\right)
p^{2}+2\kappa_{\text{tr}}\mathbf{p}^{2}\right]  }.
\end{align}

The residues in the double pole $p^{4}=0$\ are given by\textbf{\ }%
\begin{equation}
\text{Res}\left[  \tilde{\Delta}_{00}\right]  =-\frac{1}{1+\lambda
}~\ ,~\ \ \ \text{Res}\left[  \tilde{\Delta}_{0k}\right]
=0\ ,~\ \ \ \text{Res}\left[  \tilde{\Delta}_{jk}\right]  =\frac{p_{j}p_{k}%
}{\left(  \kappa_{\text{tr}}+1\right)  \mathbf{p}^{2}}.
\end{equation}
By using the current conservation condition in momentum space, $p_{\mu}%
\tilde{J}^{\mu}=0$, and the fact $p^{2}=0$, we obtain as saturation
\begin{equation}
SP=\left(  J_{0}\right)  ^{2}\frac{\lambda-\kappa_{\text{tr}}}{\left(
1+\lambda\right)  \left(  \kappa_{\text{tr}}+1\right)  },
\end{equation}
which is gauge dependent. Therefore, the pole $p^{2}=0$\ is nonphysical, once
such behavior is not compatible with the a physical pole. On the other hand,
when the saturation is evaluated for the physical pole $p^{2}=-2\kappa
_{\text{tr}}p^{2}/(1+\kappa_{\text{tr}}),$ the result is independent of the
gauge parameter.

\ Moreover, it is possible to show that if we choose $\lambda=\kappa
_{\text{tr}}$ or $\xi=1/(\kappa_{\text{tr}}+1),$ the double pole does not
appear longer in the propagator, becoming a single pole in $p^{2}.$\ So, its
own existence depends on the choice of the gauge fixing parameter. This fact
is also not compatible with a physical pole.

\section{An alternative discussion on unitarity \label{apendice2}}

In this Appendix, we perform an alternative and complementary analysis on
unitarity which confirms the previous results of this work. For the parity-odd
sector, we present a unitarity calculation for two configurations,
$\boldsymbol{\kappa}\parallel\mathbf{p}$ and $\boldsymbol{\kappa}%
\perp\mathbf{p}$. For the parity-even sector, we present a evaluation that
holds for arbitrary momentum and use the current conservation. In both
situations, without loss generality, we set $\xi=0$.

We begin discussing the parity-odd case, for the case $\mathbf{\kappa}$ is
parallel to $\mathbf{p},$ writing $\boldsymbol{\kappa}=\lambda\mathbf{p}.$
Following it, we write the terms of the propagator as
\begin{align}
\tilde{\Delta}_{00}\left(  p\right)   &  =-\frac{1}{p^{2}},\text{ }%
\tilde{\Delta}_{0j}\left(  p\right)  =\tilde{\Delta}_{j0}\left(  p\right)
=0,\\
\tilde{\Delta}_{ij}\left(  p\right)   &  =\frac{\delta_{ij}}{p^{2}+2\lambda
p_{0}\mathbf{p}^{2}}+\frac{2\lambda p_{0}p_{i}p_{j}}{p^{2}\left[
p^{2}+2\lambda p_{0}\mathbf{p}^{2}\right]  }.
\end{align}
Observing these terms, we identify only two first order poles, the nonphysical
$p^{2}=0~$and the physical $p^{2}=-2\lambda p_{0}p^{2}$.

We should now analyze the pole $p^{2}=-2\lambda p_{0}\mathbf{p}^{2},$ for
which the residues are%

\begin{equation}
\text{Res}\left[  \tilde{\Delta}_{00}\left(  p\right)  \right]  =\text{Res}%
\left[  \tilde{\Delta}_{0i}\left(  p\right)  \right]  =0,\ \text{Res}\left[
\tilde{\Delta}_{ij}\left(  p\right)  \right]  =\delta_{ij}-p_{i}%
p_{j}/\mathbf{p}^{2}.
\end{equation}
The implied saturation is $SP=\mathbf{\tilde{J}}^{2}-\left(  p_{j}\tilde
{J}^{j}\right)  ^{2}/\mathbf{p}^{2}=\mathbf{\tilde{J}}^{2}-\left(
\mathbf{p\cdot\tilde{J}}\right)  ^{2}/\mathbf{p}^{2},$ which can be written as
a positive definite expression,%
\begin{equation}
SP=\left(  \mathbf{p\times\tilde{J}}\right)  ^{2}/\mathbf{p}^{2}.
\end{equation}
This positive saturation yields unitarity preservation for the pole
$p^{2}=-2\lambda p_{0}\mathbf{p}^{2}$ in the configuration $\boldsymbol{\kappa
} \parallel\mathbf{p}.$

We now consider the case in which $\boldsymbol{\kappa}\perp\mathbf{p,}$ for
which $\boldsymbol{\kappa}\cdot\mathbf{p}=0$, $\mathbf{R}=\boldsymbol{\kappa
}^{2}\mathbf{p}^{2}$. The terms of the propagator are%
\begin{align}
\tilde{\Delta}_{00}\left(  p\right)   &  =-\frac{1}{p^{2}}+\frac
{\mathbf{p}^{4}\boldsymbol{\kappa}^{2}}{p^{4}\left[  p^{2}-\boldsymbol{\kappa
}^{2}\mathbf{p}^{2}\right]  },\\
\tilde{\Delta}_{0j}\left(  p\right)   &  =\tilde{\Delta}_{j0}\left(  p\right)
=\frac{\mathbf{p}^{2}\kappa_{j}}{p^{2}\left[  p^{2}-\boldsymbol{\kappa}%
^{2}\mathbf{p}^{2}\right]  }+\frac{p_{0}p_{j}\boldsymbol{\kappa}^{2}%
\mathbf{p}^{2}}{p^{4}\left[  p^{2}-\boldsymbol{\kappa}^{2}\mathbf{p}%
^{2}\right]  },
\end{align}%
\begin{equation}
\tilde{\Delta}_{ij}\left(  p\right)  =\frac{\delta_{ij}}{p^{2}}+\frac
{p_{0}\left(  \kappa_{i}p_{j}+\kappa_{j}p_{i}\right)  +\kappa_{i}\kappa
_{j}\mathbf{p}^{2}+p_{i}p_{j}\boldsymbol{\kappa}^{2}}{p^{2}\left[
p^{2}-\boldsymbol{\kappa}^{2}\mathbf{p}^{2}\right]  }+\frac{p_{i}%
p_{j}\boldsymbol{\kappa}^{2}\mathbf{p}^{2}\ \ }{p^{4}\left[  p^{2}%
-\boldsymbol{\kappa}^{2}\mathbf{p}^{2}\right]  }.
\end{equation}
Now, we consider the physical first order pole, $p^{2}=\kappa^{2}p^{2}$, for
which we evaluate the residues for the pole $p^{2}=\boldsymbol{\kappa}%
^{2}\mathbf{p}^{2},$%
\begin{equation}
\text{Res}\left[  \tilde{\Delta}_{00}\left(  p\right)  \right]  =\frac
{1}{\boldsymbol{\kappa}^{2}},\text{ Res}\left[  \tilde{\Delta}_{0j}\left(
p\right)  \right]  =\frac{\kappa_{j}}{\boldsymbol{\kappa}^{2}}+\frac
{p_{0}p_{j}}{\boldsymbol{\kappa}^{2}\mathbf{p}^{2}},
\end{equation}%
\begin{equation}
\text{Res}\left[  \tilde{\Delta}_{ij}\left(  p\right)  \right]  =\frac
{p_{0}\left(  \kappa_{i}p_{j}+\kappa_{j}p_{i}\right)  +\kappa_{i}\kappa
_{j}\mathbf{p}^{2}+p_{i}p_{j}\boldsymbol{\kappa}^{2}}{\boldsymbol{\kappa}%
^{2}\mathbf{p}^{2}}+\frac{p_{i}p_{j}\ }{\boldsymbol{\kappa}^{2}\mathbf{p}^{2}%
}.
\end{equation}
The associated saturation,%
\begin{equation}
SP=\frac{1}{\boldsymbol{\kappa}^{2}}\left[  \left(  \mathbf{\kappa\cdot
\tilde{J}}\right)  +\tilde{J}^{0}\boldsymbol{\kappa}^{2}\right]  ^{2},
\end{equation}
is always positive, which yields unitarity preservation. Thus, the pole
$p^{2}=\boldsymbol{\kappa}^{2}\mathbf{p}^{2}$ is a physical one.

Finally, we consider the parity-even case. In order to correctly evaluate the
saturation at this pole, we write the propagator elements as:%
\begin{align}
\tilde{\Delta}_{00} &  =-\frac{\left(  \kappa_{\text{tr}}+1\right)
p^{2}+\kappa_{\text{tr}}\mathbf{p}^{2}}{\left(  \kappa_{\text{tr}}+1\right)
p^{4}},\text{ \ }\tilde{\Delta}_{0k}=\tilde{\Delta}_{0k}=-\frac{\kappa
_{\text{tr}}p_{0}p_{k}}{\left(  \kappa_{\text{tr}}+1\right)  p^{4}%
},\label{res1e}\\
\tilde{\Delta}_{jk} &  =\frac{\delta_{jk}}{\left[  \left(  \kappa_{\text{tr}%
}+1\right)  p^{2}+2\kappa_{\text{tr}}\mathbf{p}^{2}\right]  }+\frac
{\kappa_{\text{tr}}p_{j}p_{k}\left[  \left(  1-\kappa_{\text{tr}}\right)
p^{2}-2\kappa_{\text{tr}}\mathbf{p}^{2}\right]  }{\left(  \kappa_{\text{tr}%
}+1\right)  p^{4}\left[  \left(  \kappa_{\text{tr}}+1\right)  p^{2}%
+2\kappa_{\text{tr}}\mathbf{p}^{2}\right]  }.\label{res2e}%
\end{align}
We observe a nonphysical double pole in $p^{2}=0$\ and a physical single pole
in $p^{2}=-2\kappa_{\text{tr}}p^{2}/\left(  \kappa_{\text{tr}}+1\right)  $. We
consider now only the physical pole $[\left(  \kappa_{\text{tr}}+1\right)
p^{2}+2\kappa_{\text{tr}}\mathbf{p}^{2}]=0$, for which the residue of
propagator (\ref{res1e},\ref{res2e}) is Res$\left(  \tilde{\Delta}%
_{00}\right)  =0,$ Res$\left(  \tilde{\Delta}_{0k}\right)  =$Res$\left(
\tilde{\Delta}_{k0}\right)  =0,$ Res$\left(  \tilde{\Delta}_{jk}\right)
=\left(  \delta_{jk}-p_{j}p_{k}/\mathbf{p}^{2}\right)  /(\kappa_{\text{tr}%
}+1)$. The saturation of the residue propagator with the current is
\begin{equation}
SP=\frac{1}{\kappa_{\text{tr}}+1}\frac{\left(  \mathbf{p}\times\mathbf{J}%
\right)  ^{2}}{\mathbf{p}^{2}}.
\end{equation}
This implies a positive saturation whenever $\kappa_{\text{tr}}>-1.$ Thus, we
see that these calculations confirm our previous results.

\begin{acknowledgments}
The authors are grateful for FAPEMA, CAPES and CNPq (Brazilian research
agencies) for invaluable financial support. The authors also thank Jose A.
Helayel-Neto, Ralf Lehnart and Matthew Mewes for relevant comments on this work.
\end{acknowledgments}


\begin{thebibliography}{99}                                                                                               %
\bibitem {Colladay}D. Colladay and V. A. Kostelecky, \textit{Phys. Rev. }D
\textbf{55}, 6760 (1997); D. Colladay and V. A. Kostelecky, \textit{Phys. Rev.
}D \textbf{58}, 116002 (1998); S. R. Coleman and S. L. Glashow, \textit{Phys.
Rev}. D\textbf{\ 59}, 116008 (1999).

\bibitem {Samuel}V. A. Kostelecky and S. Samuel, \textit{Phys. Rev. Lett}.
\textbf{63}, 224 (1989); \textit{Phys. Rev. Lett}. \textbf{66}, 1811 (1991);
\textit{Phys. Rev. }D\textbf{\ 39}, 683 (1989); \textit{Phys. Rev.
}D\textbf{\ 40}, 1886 (1989), V. A. Kostelecky and R. Potting, \textit{Nucl.
Phys.} B\textbf{\ 359}, 545 (1991); Phys. Lett. B \textbf{381}, 89 (1996); V.
A. Kostelecky and R. Potting, \textit{Phys. Rev. }D\textbf{\ 51}, 3923 (1995).

\bibitem {General}N.M. Barraz, Jr., J.M. Fonseca, W.A. Moura-Melo, and J.A.
Helayel-Neto, Phys. Rev. D\textbf{76}, 027701 (2007); H. Belich , J.L. Boldo,
L.P. Colatto, J.A. Helayel-Neto, A.L.M.A. Nogueira, Phys.Rev. D \textbf{68},
065030 (2003); A.P. Baeta Scarpelli, H. Belich, J.L. Boldo, L.P. Colatto, J.A.
Helayel-Neto, A.L.M.A. Nogueira, Nucl. Phys. Proc. Suppl.127, 105-109 (2004);
M.N. Barreto, D. Bazeia, and R. Menezes, Phys. Rev. D \textbf{73}, 065015
(2006); M. B. Cantcheff, Eur. Phys. J. C \textbf{46}, 247 (2006); M. B.
Cantcheff, C.F.L. Godinho, A.P. Baeta Scarpelli, J.A. Helay\"{e}l-Neto, Phys.
Rev. D \textbf{68}, 065025 (2003); H. Belich, T. Costa-Soares, J.A.
Helayel-Neto M.T.D. Orlando, R.C. Paschoal, Phys. Lett. A \textbf{370}, 126
(2007); H. Belich, L.P. Colatto, T. Costa-Soares, J.A. Helayel-Neto, M.T.D.
Orlando, Eur. Phys. J. C \textbf{62}, 425 (2009); F.A. Brito, L.S. Grigorio,
M.S. Guimaraes, E. Passos, C. Wotzasek, Phys.Rev. D 78, 125023 (2008); M. A.
Anacleto, C. Furtado, J. R. Nascimento, A. Yu. Petrov, Phys. Rev. D
\textbf{78}, 065014 (2008); B. Charneski, M. Gomes, T. Mariz, J. R.
Nascimento, A. J. da Silva, Phys. Rev. D \textbf{79}, 065007 (2009).

\bibitem {Fermion}B. Altschul, Phys. Rev. D \textbf{70}, 056005 (2004); G. M.
Shore, Nucl. Phys. B \textbf{717}, 86 (2005); \ D. Colladay and V. A.
Kostelecky, Phys. Lett. B \textbf{511}, 209 (2001); M. M. Ferreira Jr, Phys.
Rev. D \textbf{70,} 045013 (2004); M. M. Ferreira Jr, Phys. Rev. D
\textbf{71,} 045003 (2005); M. M. Ferreira Jr and M. S. Tavares, Int. J. Mod.
Phys. \textbf{A} 22, \ 1685 (2007); H. Belich, T. Costa-Soares, M.M. Ferreira
Jr., J. A. Helay\"{e}l-Neto, and F. M. O. Moucherek, Phys. Rev. D \textbf{74},
065009 (2006); O. G. Kharlanov and V. Ch. Zhukovsky, J. Math. Phys.
\textbf{48}, 092302 (2007); R. Lehnert, Phys. Rev. D \textbf{68}, 085003
(2003); V.A. Kostelecky and C. D. Lane, J. Math. Phys. \textbf{40}, 6245
(1999); R. Lehnert, J. Math. Phys. \textbf{45}, 3399 (2004).

\bibitem {Lehnert}V. A. Kostelecky and R. Lehnert, Phys. Rev. D\textbf{\ 63,
}065008 (2001)\textbf{.}

\bibitem {Tests}S.R. Coleman and S.L. Glashow, \textit{Phys. Rev}.
D\textbf{\ 59}, 116008 (1999); V. A. Kostelecky and M. Mewes, Phys. Rev. Lett.
\textbf{87}, 251304 (2001); V. A. Kostelecky and M. Mewes, Phys. Rev.
D\ \textbf{66}, 056005 (2002); J. Lipa, J.A. Nissen, S.Wang, D. A. Stricker,
D. Avaloff, Phys. Rev. Lett. \textbf{90}, 060403 (2003); R. Bluhm, V.A.
Kostelecky, and N. Russell, Phys. Rev. Lett. \textbf{79}, 1432 (1997); R.
Bluhm, V.A. Kostelecky, and N. Russell, Phys. Rev. D \textbf{57}, 3932 (1998);
R. Bluhm, V.A. Kostelecky, C. D. Lane, and N. Russell, Phys. Rev. Lett.
\textbf{88}, 090801 (2002); R. Bluhm and V.A. Kostelecky, Phys. Rev. Lett.
\textbf{84}, 1381 (2000); R. Bluhm, R. Bluhm, V.A. Kostelecky, and C. D. Lane,
Phys. Rev. Lett. \textbf{84}, 1098 (2000); R. Bluhm, V.A. Kostelecky, N.
Russell, Phys. Rev. Lett. \textbf{82}, 2254 (1999); V.A. Kostelecky and C. D.
Lane, Phys. Rev. D \textbf{60}, 116010 (1999).

\bibitem {Jackiw}S.M. Carroll, G.B. Field and R. Jackiw, Phys. Rev.\textit{\ }%
D \textbf{41}, 1231 (1990).

\bibitem {Adam}C. Adam and F. R. Klinkhamer, Nucl. Phys.\textit{\ }B
\textbf{607}, 247 (2001); C. Adam and F. R. Klinkhamer, Nucl. Phys.\textit{\ }%
B \textbf{657}, 214 (2003).

\bibitem {Soldati}A.A. Andrianov and R. Soldati, Phys. Rev. D \textbf{51},
5961 (1995); Phys. Lett. B \textbf{435}, 449 (1998); A.A. Andrianov, R.
Soldati and L. Sorbo, Phys. Rev. D \textbf{59}, 025002 (1998).

\bibitem {Higgs}A. P. Baeta Scarpelli, H. Belich, J. L. Boldo, J.A.
Helayel-Neto, Phys. Rev. D\textit{\ }\textbf{67}, 085021 (2003).

\bibitem {Belich}H. Belich, M.M. Ferreira Jr., J.A. Helayel-Neto, M.T.D.
Orlando, Phys. Rev. D \textbf{67},125011 (2003); Erratum-ibid., Phys. Rev.
\textbf{D }69, 109903 (2004).

\bibitem {Casana}R. Casana, M.M. Ferreira Jr, C. E. H. Santos, Phys. Rev.
\textbf{D 78}, 025030 (2008).

\bibitem {Cerenkov1}R. Lehnert and R. Potting, Phys. Rev. Lett. \textbf{93},
110402 (2004); R. Lehnert and R. Potting, Phys. Rev. D \textbf{70}, 125010
(2004); B. Altschul, Phys. Rev. D \textbf{75}, 105003 (2007); C. Kaufhold and
F.R. Klinkhamer, Nucl. Phys. B \textbf{734, 1 }(2006).

\bibitem {CBR}R. Casana, M. M. Ferreira Jr. and J. S. Rodrigues, Phys. Rev. D
\textbf{78}, 125013 (2008).

\bibitem {winder}J. M. Fonseca, A. H. Gomes, W. A. Moura-Melo, Phys. Lett. B
\textbf{671}, 280 (2009).

\bibitem {CBR2}R. Casana, M. M. Ferreira Jr., J. S. Rodrigues, Madson R.O.
Silva, Phys. Rev. D \textbf{80}, 085026 (2009).

\bibitem {Radio}R. Jackiw and V. A. Kosteleck\'{y}, Phys. Rev. Lett.
\textbf{82}, 3572 (1999); J. M. Chung and B. K. Chung Phys. Rev.
D\textbf{\ 63}, 105015 (2001); J.M. Chung, Phys.Rev. D\ \textbf{60}, 127901
(1999); G. Bonneau, Nucl.Phys. B\ \textbf{593}, 398 (2001); M. Perez-Victoria,
Phys. Rev. Lett. \textbf{83}, 2518 (1999); M. Perez-Victoria, J. High. Energy
Phys. \textbf{\ 0104}, (2001) 032; O.A. Battistel and G. Dallabona, Nucl.
Phys. B \textbf{610}, 316 (2001); O.A. Battistel and G. Dallabona, J. Phys. G
\textbf{28}, L23 (2002); J. Phys. G \textbf{27}, L53 (2001); A. P. B.
Scarpelli, M. Sampaio, M. C. Nemes, and B. Hiller, Phys. Rev. D\ \textbf{64},
046013 (2001); T. Mariz, J.R. Nascimento, E. Passos, R.F. Ribeiro and F.A.
Brito, J. High. Energy Phys. \textbf{0510} (2005) 019; J. R. Nascimento, E.
Passos, A. Yu. Petrov, F. A. Brito, J. High. Energy Phys. \textbf{0706},
(2007) 016; B. Altschul, Phys. Rev. D \textbf{70}, 101701 (2004); A.P.B.
Scarpelli, M. Sampaio, M.C. Nemes, B. Hiller, Eur. Phys. J. C \textbf{56}, 571 (2008).

\bibitem {KM1}V. A. Kostelecky and M. Mewes, Phys. Rev. Lett. \textbf{87},
251304 (2001).

\bibitem {KM2}V. A. Kostelecky and M. Mewes, Phys. Rev. D\textbf{\ 66}, 056005 (2002).

\bibitem {KM3}V. A. Kostelecky and M. Mewes, Phys. Rev. Lett. \textbf{97},
140401 (2006); Astrophys. J. Lett. \textbf{689}, L1 (2008).

\bibitem {Bailey}Q. G. Bailey and V. A. Kostelecky, Phys. Rev. D \textbf{70},
076006 (2004).

\bibitem {Kostelec}V. A. Kostelecky and M. Mewes, Phys. Rev.\textit{\ }D
\textbf{80}, 015020\ (2009).

\bibitem {Cherenkov2}B. Altschul, Nucl. Phys. B\textbf{\ 796}, 262 (2008); B.
Altschul, Phys. Rev. Lett. \textbf{98}, 041603 (2007); C. Kaufhold and F.R.
Klinkhamer, Phys. Rev. D \textbf{76}, 025024 (2007)\textbf{. }

\bibitem {Klink2}F.R. Klinkhamer and M. Risse, Phys. Rev. D \textbf{77},
016002 (2008); F.R. Klinkhamer and M. Risse, Phys. Rev. D \textbf{77}, 117901
(A) (2008).

\bibitem {Klink3}R. Klinkhamer and M. Schreck, Phys. Rev. D \textbf{78},
085026 (2008).

\bibitem {Interac1}V. A. Kostelecky and A.G.M. Pickering, Phys. Rev. Lett.
\textbf{91}, 031801 (2003); B. Altschul, Phys.Rev. D \textbf{70, }056005 (2004).

\bibitem {Interac2}C.D. Carone, M. Sher, and M. \ Vanderhaeghen, Phys. Rev. D
\textbf{74}, 077901 (2006); B. Altschul, Phys. Rev. D \textbf{79}, 016004 (2009).

\bibitem {Interac3}M.A. Hohensee, R. Lehnert, D. F. Phillips, R. L. Walsworth,
Phys. Rev. D \textbf{80}, 036010(2009); M.A. Hohensee, R. Lehnert, D. F.
Phillips, R. L. Walsworth, Phys. Rev. Lett. \textbf{102}, 170402 (2009); B.
Altschul, Phys. Rev. D \textbf{80}, 091901(R) (2009).

\bibitem {Manojr1}R. Casana, M.M. Ferreira Jr, C. E. H. Santos, Phys. Rev.
D\textbf{\ 78}, 105014 (2008).

\bibitem {Paulo}R. Casana, M.M. Ferreira Jr, A. R. Gomes, P. R. D. Pinheiro,
Eur. Phys. J. C \textbf{62}, 573 (2009).

\bibitem {Kob}A. Kobakhidze and B.H.J. McKellar, Phys. Rev. D \textbf{76},
093004 (2007).

\bibitem {Sexl}R. U. Sexl and H.K. Urbantke, "Relativity, Groups, Particles:
special relativity and relativistic symmetry in field and particle physics",
Springer-Verlag, New York (1992).

\bibitem {Veltman}M. Veltman, "Quantum Theory of Gravitation", in Methods in
Field Theory Ed.by R. Bailian and J. Zinn-Justin, North-Holland Publising
Company and World Scientific Publising Co Ltd, Singapore, 1981.

\bibitem {Unitarity}A. P. Baeta Scarpelli, H. Belich, J. L. Boldo, and J. A.
Helayel-Neto, Phys. Rev. D \textbf{67}, 085021 (2003); A. P. Baeta Scarpelli
and J. A. Helayel-Neto, Phys. Rev. D \textbf{73}, 105020 (2006).
\end{thebibliography}
\end{document}